\documentclass[nohyper,12pt,letterpaper]{JHEP3}
\usepackage{epsfig,youngtab}

\author{Robert de Mello Koch$^{1,2}$ and Jeff Murugan$^{2,3}$\\
\qquad \\
$^{1}$ National Institute for Theoretical Physics,\\
Department of Physics and Centre for Theoretical Physics,\\ 
University of the Witwatersrand,\\ 
Wits, 2050,\\ 
South Africa\\
\qquad\\
$^{2}$National Institute for Theoretical Physics,\\
Stellenbosch,\\
South Africa\\
\qquad\\
$^{3}$Cosmology and Gravity Group,\\
Department of Mathematics and Applied Mathematics,\\
University of Cape Town,\\
Private Bag, Rondebosch, 7700,\\
South Africa\\

E-mail: \email{robert@neo.phys.wits.ac.za,jeff@nassp.uct.ac.za}}

\abstract{
We give an introductory account of the AdS/CFT correspondence in the ${1\over 2}$-BPS sector of ${\cal N}=4$ super Yang-Mills theory.
Six of the dimensions of the string theory are emergent in the Yang-Mills theory. In this
article we suggest how these dimensions and local physics in these dimensions emerge. 
The discussion is aimed at non-experts.}

\preprint{WITS-CTP-048, UCT-CGG-251109}

\title{Emergent Spacetime}


\def \Tr{\mbox{Tr\,}}

\begin{document}

\section{Introduction}

The problem of quantizing gravity has proved to be a difficult one. To solve this problem, 
it seems to be necessary to answer the question ``What is spacetime?'' This challenges the most 
basic assumptions we are used to making; a radical new idea may be needed. Further, the hope 
of any guidance from experiment seems to be out of the question. One might conclude that the 
situation is hopeless. Drawing on recent insights from the AdS/CFT correspondence, we are 
nonetheless, optimistic.

The AdS/CFT correspondence\cite{Maldacena:1997re} claims an exact equality between ${\cal N}=4$ super Yang-Mills theory
in flat $3+1$ dimensional Minkowski spacetime and type IIB string theory on an asymptotically 
AdS$_5\times$S$^5$ background. 
Type IIB string theory is a theory of closed strings; at least within string
perturbation theory, theories of closed strings provide a consistent UV 
completion of gravity.
The fact that such an equality exists is highly unexpected and 
nontrivial, and (as we will try to convince the reader) can be used to gain insight into the 
nature of spacetime. George Ellis opened the ``{\it Foundations of Space and Time}" workshop 
by holding up two fingers and asking ``are there
an infinite or a finite number of places a particle could occupy between my fingers?'' We don't
know the answer to George's question. However, we hope to convince the reader that the AdS/CFT
correspondence provides a detailed and concrete framework within which this question can be 
tackled.

We know how to formulate ${\cal N}=4$ super Yang-Mills theory as a path integral. We do not yet
understand how to formulate quantum gravity in asymptotically AdS$_5\times$S$^5$ backgrounds.
Its seems somewhat natural then, to use the ${\cal N}=4$ super Yang-Mills theory as a definition 
for quantum gravity in asymptotically AdS$_5\times$S$^5$ backgrounds. The puzzle then becomes one
of translating and interpreting the quantum field theory, as a quantum theory of gravity. Conceptually
this is challenging and we do not have any simple arguments that would explain why a higher dimensional
gravity theory is encoded in the dynamics of a quantum field theory. Technically its tough too. The 
relation between the radius of the AdS space (measured in units of the string length $l_s$) and the 
't Hooft coupling\footnote{Recall that by suitably rescaling the fields one can arrange things so that
all $g_{YM}^2$ dependence factors out as an overall ${1\over g_{YM}^2}$ factor in front of the action. 
It is then clear that $g_{YM}^2$ plays the role of $\hbar$ for the quantum field theory.} 
($\lambda=g_{YM}^2 N$)
$$ {R^4_{\rm AdS}\over l_s^4}=\lambda$$
shows that in the limit of small curvatures (where we could have hoped to recognize a familiar description
of geometry) the field theory is strongly coupled and hence we do not know how to do the relevant field 
theory calculations. Conversely, if we compute things perturbatively in the field theory, we are studying
the small $\lambda$ limit where curvature corrections are important and our usual notions of geometry are
probably not useful.

Fortunately, there is a way to proceed. Thanks to the large amount of supersymmetry enjoyed by the theory,
there do exist quantities that are protected from corrections. If one chooses carefully, these quantities can be
computed at weak coupling and the result can then be extrapolated to strong coupling. The most protected
states of the theory, preserving half of the maximal amount of supersymmetry, are called the
${1\over 2}$-BPS sector. This is the laboratory in which we will work.

In section 2 we will give some arguments for the simplicity of the ${1\over 2}$-BPS sector. In section 3
we will explain how the dictionary between the gauge theory and the gravity theory is organized - its
organized according to the ${\cal R}$-charge\footnote{The ${\cal R}$-charge is a conserved charge
associated specifically with supersymmetric theories. Recall that an internal symmetry is one whose generators
commute with all of the spacetime symmetry generators. An ${\cal R}$ symmetry is one whose generators commute
with all of the bosonic spacetime symmetries but fail to commute with the fermionic supercharges.} of the 
operators of the field theory. Section 4 introduces a set of variables, the Schur polynomials, which 
provides a beautiful organization of the degrees of freedom of the theory. In sections 5, 6 and 7
we explain how to describe gravitons, strings and branes of the string theory using the field theory
language, and in section 8 we explain how new backgrounds (the so called LLM geometries) arise. Section
9 is reserved for discussion.

There are a number of papers that have significantly influenced our point of view and have had an impact on
our research. Among these we mention 
\cite{Das:1990kaa,Corley:2001zk,Berenstein:2004kk,Lin:2004nb,Balasubramanian:2005mg,Berenstein,Balasubramanian:2004nb,sanjaye}. 

\section{Simplicity of the ${1\over 2}$-BPS Sector}

We study ${\cal N}=4$ super Yang-Mills theory on $R\times S^3$. 
The field content of ${\cal N}=4$ super Yang-Mills theory includes 6 Hermitian scalars transforming in the adjoint
of the gauge group. We group the six real scalars into three complex fields as follows
$$ Z=\phi_1+i\phi_2,\qquad Y=\phi_3+i\phi_4,\qquad X=\phi_5+i\phi_6\, . $$
The half BPS chiral primary operators we focus on can be built from a single complex combination (we use $Z$ in what follows).
Using a total of $n$ $Z$s, there is a distinct operator for each partition of $n$. 
Given the partition with parts $\{ n_i\} $, the corresponding operator is
$\prod_i \Tr (Z^{n_i})$.
There is a one to one correspondence between 
these operators and half BPS representations of $\mathcal{R}$-charge $n$\cite{Corley:2001zk}.

A beautiful argument, due to Berenstein\cite{Berenstein:2004kk}, demonstrates 
the simplicity of the ${1\over 2}$-BPS sector\footnote{See \cite{Berenstein:2002jq}
for closely related ideas.}. Consider a time slicing of AdS$_5\times$S$^5$, which gives the Hamiltonian
$$ H={(\Delta -J)+\epsilon \Delta\over \epsilon} \, ,$$
where $\Delta$ is the dilatation operator and $J$ is the ${\cal R}$-charge under which $Z$ has one unit of charge. In the limit
$\epsilon\to 0$ any state with $\Delta - J>0$ will have a huge energy and hence will decouple from the low energy theory. This
procedure decouples (a subspace) of the ${1\over 2}$-BPS states of ${\cal N}=4$ super Yang-Mills theory. These low lying states
are protected by supersymmetry and will not be lifted from zero energy by interactions. In what follows, we assume that, even in
the presence of interactions, these states remain decoupled (which amounts to assuming that interactions do not make any of
the heavy states light). The complex scalar $Z$ can be decomposed into partial waves on the
$S^3$. The $s$-wave is simply a singlet under the $SO(4)$ symmetry of the $S^3$ on which the field theory is defined. The higher
spherical harmonics have a greater energy and hence are among the states that decouple. We thus come to the remarkable conclusion
that the limit we study is described exactly by the quantum mechanics of a single complex matrix.
The action of ${\cal N}=4$ super Yang-Mills theory on $R\times S^3$ includes a mass term which arises from 
conformal coupling to the metric of $S^3$. With a convenient normalization of the action, the free field 
theory propagators are
$$ \langle Z^\dagger_{ij}(t)Z_{kl}(t)\rangle = \delta_{il}\delta_{jk} = 
\langle Y^\dagger_{ij}(t)Y_{kl}(t)\rangle = 
\langle X^\dagger_{ij}(t)X_{kl}(t)\rangle .$$
As long as one restricts attention to traces involving only $Z$ or only $Z^\dagger$, it is possible to express the theory in
terms of the eigenvalues of $Z$. The change of variables entailed in going from $Z$ to the eigenvalues of $Z$ induces a non-trivial
Jacobian - the Van der Monde determinant. The net effect of this Jacobian is accounted for by treating the eigenvalues as 
fermions\cite{Brezin:1977sv}.
Consequently, one obtains the dynamics of $N$ non-interacting fermions in an external harmonic oscillator potential.

{\vskip 0.5cm}

\noindent
{\sl {\bf Key idea:} The ${1\over 2}$-BPS sector of ${\cal N}=4$ super Yang-Mills theory
is described exactly by the holomorphic sector of the 
quantum mechanics of a single complex matrix, which itself is 
equivalent to the dynamics of $N$ free
fermions.}

\section{Dictionary}

The ${1\over 2}$-BPS sector of type IIB string theory on AdS$_5\times$S$^5$ contains gravitons, membranes and strings. Apparently
all of these objects are captured by the matrix quantum mechanics of the previous section. To see that this is indeed plausible,
recall that as the ${\cal R}$-charge ($J$) of an operator in the ${\cal N}=4$ super Yang-Mills theory is changed, 
its interpretation in the dual quantum gravity changes. This is a consequence of the Myers 
effect\cite{Myers:1999ps}: the background we are studying has a non-zero RR five form field strength switched on. This flux couples to
D3 branes. Gravitons carry a D3 dipole charge and are hence polarized by the background flux\cite{McGreevy:2000cw}.
As we increase $J$, the coupling to the background RR five form flux increases 
and the graviton expands. It puffs out to a radius
$$ R=\sqrt{J\over N}R_{\rm AdS},\qquad{\rm where}\qquad R_{\rm AdS}^2=\sqrt{g_{YM}^2 N} \alpha'\, . $$
We will consider the limit that $N$ is very large with $g_{YM}^2$ fixed and very small. Since the string coupling $g_s=g_{YM}^2$,
this is the weak string coupling and small curvature limit in which we expect to be able to recognize the familiar objects of perturbative
string theory.
For $J\sim O(1)$ the operator is dual to an object of zero size in string units, that is, a point-like 
graviton\cite{Maldacena:1997re}.
For $J\sim O(\sqrt{N})$ the operator is dual to an object of fixed size in string units - this is a string\cite{Berenstein:2002jq}.
For $J\sim O(N)$ the operator is dual to an object whose size is of the order of $R_{\rm AdS}$ - as argued in 
\cite{Balasubramanian:2001nh,Corley:2001zk} these are the giant gravitons of \cite{McGreevy:2000cw}.
Finally, consider $J\sim O(N^2)$. Naively, the size of these objects diverge, even when
measured in units with $R_{\rm AdS}=1$. We will see that this divergence is simply an indication 
that these operators do not have an interpretation
in terms of a new object in AdS$_5\times$S$^5$: these operators correspond to new backgrounds \cite{Lin:2004nb,Balasubramanian:2005mg}.

The original ${\cal N}=4$ super Yang-Mills theory is defined on Minkowski space. After Wick rotating (to four dimensional Euclidean space)
and performing a conformal
transformation, we obtain ${\cal N}=4$ super Yang-Mills theory on $R\times S^3$. Operators of the theory on four dimensional Euclidean
space are in one-to-one correspondence with states of the theory on $R\times S^3$, by the usual operator-state correspondence
available for any conformal field theory. $R\times S^3$ is the boundary of AdS$_5\times$S$^5$
in global coordinates. It is natural to identify this boundary with the space on which the field theory lives. Taken together, we obtain
a map between operators of the original ${\cal N}=4$ super Yang-Mills theory on Minkowski space and states of the string theory. For
this reason we will often talk of ``matching operators to states'' and we will often be 
able to read the inner product between two states from
a correlation function of the corresponding operators.

The symmetries of the ${\cal N}=4$ super Yang-Mills theory match the isometries present in the dual AdS$_5\times$S$^5$ background.
When trying to match a specific operator to a specific state, it is useful to match the labels provided by these symmetries on both
sides of the correspondence. Reasoning in this way, it is possible to argue that scaling dimensions in the field theory correspond
to energies in the dual string theory and ${\cal R}$-charge in the field theory corresponds to angular momentum in the string theory.

{\vskip 0.5cm}

\noindent
{\sl {\bf Key idea:} The dictionary between the ${1\over 2}$-BPS sector of ${\cal N}=4$ super Yang-Mills theory
and type IIB string theory on AdS$_5\times$S$^5$
is organized according to ${\cal R}$-charge. When trying to match a specific operator to a specific state, it is useful to match
scaling dimensions (${\cal R}$-charge) in the field theory to energies (angular momentum) in the dual string theory.}

\section{Organizing the degrees of freedom of a matrix model}

In the previous section we have seen that, in order to capture all of the objects in the spectrum of the dual string theory,
it is necessary to consider all possible values of the ${\cal R}$-charge. This is a complicated problem since
the usual simplifications of the large $N$ limit are no longer present, as we now explain.

First consider the case that $n=O(1)$. A suitable basis is provided by single trace operators, $\Tr (Z^n)$.
Normalize the operators with factors of $N$ so that they have an $O(1)$ two point function at large $N$
$$ {\cal O}_n={\Tr (Z^n)\over N^{n\over 2}}\, . $$
In this case we obviously have
$$ \langle {\cal O}_n{\cal O}_m^\dagger\rangle \propto\delta_{mn}\, ,$$
because the two operators have a different ${\cal R}$-charge if $m\ne n$. Consider next the correlator
$$\langle {\cal O}_p{\cal O}_n{\cal O}_{n+p}^\dagger\rangle\sim {1\over N}\, .$$
The total ${\cal R}$-charge of this three point function is zero, so it is not forced to vanish. 
However, recall that in the large $N$ limit the expectation values of observables factorize, which is equivalent
to that statement that disconnected diagrams dominate. The leading (disconnected) contributions to the above
correlator vanish and hence this correlator is suppressed in the large $N$ limit by the usual arguments.  
To explain why this correlator vanished, we can identify the number of traces with particle number 
in which case the vanishing of the above correlator is the statement that
although a two particle state with gravitons of ${\cal R}$-charge $p$ and $n$ has 
the same ${\cal R}$-charge as a single
particle state with ${\cal R}$-charge $p+n$, the two states are orthogonal. 
The fact that the two states have a different
particle number {\it explains} why their overlap is zero.
Consequently, the weakly interacting supergravity Fock
space is clearly visible in the dual gauge theory. There is a combinatoric coefficient on the right hand side of the above correlator
which counts the number of Wick contractions.

For $n=O(N)$, the usual ${1\over N}$ suppression of non-planar diagrams is compensated by huge combinatoric factors\footnote{The number
of Wick contractions explodes as more and more operators are included in each 
trace\cite{Balasubramanian:2001nh}.}, so that operators composed of a product of a 
different number of traces, are no longer orthogonal. Clearly then, the gravity states dual to single trace operators are no longer
orthogonal and hence there is no reason to expect that they will have a natural physical interpretation. The fact that these states are
no longer orthogonal has a very natural explanation in the dual string theory. Recall that the dimension of the operator maps into the
energy of the dual state. Thus, by considering operators of a very large dimension we are talking about very heavy objects in the dual
string theory. As we increase the mass of the objects we study, we turn the gravitational interactions on and consequently the states
that were orthogonal when there was no interaction, are no longer orthogonal.

Ideally one needs a new basis in 
which the two point functions are again diagonal. Corley, Jevicki and Ramgoolam have demonstrated that the Schur polynomials (in the zero 
coupling limit) do indeed diagonalize the two point function\cite{Corley:2001zk} for the theory with gauge group $U(N)$. 
The Schur polynomial is defined by
\begin{equation}
\chi_R (Z)={1\over n!}\sum_{\sigma\in S_n}\chi_R (\sigma )\Tr (\sigma Z^{\otimes n}),
\label{Schur}
\end{equation}
$$\Tr (\sigma Z^{\otimes n}) =Z^{i_1}_{i_{\sigma (1)}}Z^{i_2}_{i_{\sigma (2)}}\cdots
Z^{i_{n-1}}_{i_{\sigma (n-1)}}Z^{i_n}_{i_{\sigma (n)}}.$$
The Schur polynomial label $R$ can be thought of as a Young diagram which has $n$ boxes. $\chi_R (\sigma )$ is the 
character of $\sigma\in S_n$ in representation $R$. For an extension of these results to the case of 
gauge group $SU(N)$ see \cite{related}.
\noindent
\begin{center}
  \fbox{
    \begin{minipage}[c]{14cm}
      \small{
        {\vskip 0.05cm}
        \center{\bf Schur Calculus}\\
        \noindent
        \flushleft{
        The dynamical content of any quantum  
        theory is encoded in its correlation functions.
        Focusing on the $\frac{1}{2}-$BPS
        sector, Schur polynomials provide an excellent set of variables to probe different aspects
        of the dual string theory since (i) in the free field limit, the two-point functions of Schur
        polynomials are known exactly and (ii) they satisfy a nice product rule that can be used to
        collapse any {\it product} of Schur polynomials into a {\it sum} of polynomials. This 
        product rule follows as a consequence of the Schur-Weyl duality between unitary 
        groups and
        symmetric groups. As a consequence of the duality, Schur polynomials
        $\chi_R(U)$,
        when evaluated on an element $U\in U(N)$ give the 
        character of $U$ in the irreducible representation $R$. For any two such
        irreducible representations, $R$ and $S$, it is well known that $R\otimes S = 
        \oplus_{T}f_{RS;T}T$ where the $f_{RS;T}$ are known as Littlewood-Richardson 
        numbers. With the interpretation of the Schur polynomials as characters, it follows 
        immediately that}
        \begin{eqnarray}
          \chi_{R}(Z)\chi_{S}(Z) = \sum_{T}f_{RS;T}\chi_{T}(Z)\nonumber
        \end{eqnarray}
        \flushleft{
          One immediate repercussion of this is that the {\it exact} computation of any 
          multi-point extremal correlator of Schur polynomials can be collapsed down to 
          an evaluation of two-point correlators\cite{Brown:2006zk}
        }.
      } 
    \end{minipage}
  }
\end{center}
There is a very natural connection between the free fermion description of section 2 and the 
Schur polynomials\cite{Corley:2001zk,Berenstein:2004kk}. The Schur polynomials
are labeled by Young diagrams, which can be specified by giving a list of $N$ integers, $r_i$, which count the number of boxes in
the $i^{th}$ row of the Young diagram. The fermion wave function can be described by specifying the $N$ occupied energy levels $E_i$,
which is again a list of $N$ integers. Detailed computations show that the Schur polynomials coincide with the $N$-fermion wave functions
as long as we identify (see \cite{Corley:2001zk,Berenstein:2004kk} for details)
$$ E_i=N+i+r_i\, .$$
Thus, the Schur polynomial basis coincides with the free fermion basis.

Although the huge simplifications discussed above do not survive when one goes beyond the ${1\over 2}$-BPS sector, it is possible
to write down more general bases which continue to diagonalize the two point 
function\cite{Brown:2007xh,Kimura:2007wy,Bhattacharyya:2008rb,organize}. 
These techniques were developed using
crucial lessons\cite{Corley:2001zk,Corley:2002mj} gained from the ${1\over 2}$-BPS sector. 
We will have more to say about these more general bases in the sections to come, since they are relevant for
describing nearly supersymmetric states and hence they suffer only mild corrections.

{\vskip 0.5cm}

\noindent
{\sl {\bf Key idea:} The basis of the ${1\over 2}$-BPS sector of ${\cal N}=4$ super Yang-Mills theory provided by the Schur polynomials
diagonalizes the free two point function for any value of the ${\cal R}$-charge. At large ${\cal R}$-charge it will thus replace the
trace basis, which now fails to diagonalize the free two point function.}

\section{Gravitons}

In this section we will focus on that portion of the AdS/CFT dictionary that concerns operators with an ${\cal R}$-charge of $O(1)$.
In this case, as explained above, the trace basis is perfectly acceptable and so we take $\mathcal{O}_{n}=\Tr (Z^n)$.
We expect that these operators are dual to gravitons. In fact, this can be checked in detail, as we now explain.

The AdS/CFT correspondence claims that for every bulk field $\Phi$ in the gravitational description, there is a corresponding gauge
invariant operator $O_\phi$. 
Asymptotically AdS spaces have a boundary at spatial infinity and one needs to impose appropriate boundary
conditions there. As a result, the partition function of the bulk theory is a functional of these boundary conditions. The boundary
values of the fields are identified with sources that couple to the dual operator so that the gravitational partition function
(the next formulas are schematic)
$$ Z_{gravity}[\phi_0]\equiv \int_{\Phi|_{\partial {\rm (AdS)}}=\phi_0}
\!\!\!\!\!\!\!\!\!\!\!\!\!\!\!\!\!\!  
\, D\Phi \,\, e^{-S} $$
is identified with the generating functional of correlation functions in the quantum field theory\footnote{The right
hand side of this relation suffers from the usual UV divergences present in any quantum field theory, and hence needs
to be renormalized. The left hand side suffers from IR divergences and hence also requires renormalization. The details
of this renormalization has been worked out in \cite{Skenderis:2002wp,Skenderis:2006uy}.}
$$ Z_{gravity}[\phi_0] = \left\langle e^{-\int \phi_0 O_\phi}\right\rangle_{\rm QFT} \, .$$
Since the gravitons are meant to be dual to operators with ${\cal R}$-charge of $O(1)$, and since graviton dynamics is captured
by the supergravity approximation to the complete string theory, using the above relation we should be able to compute the correlation
functions of the $O_n$ at strong coupling and at large $N$. Further, since these operators enjoy some protection against corrections 
by virtue of their supersymmetry, we may be optimistic that the strong coupling and weak coupling results will agree. The computation 
has been performed and the agreement is perfect\cite{Lee:1998bxa}.

{\vskip 0.5cm}

\noindent
{\sl {\bf Key idea:} The identification of gravitons with operators of ${\cal R}$-charge of O(1) can be checked by using the AdS/CFT
correspondence. The agreement is perfect.}

\section{Strings}

We now move on to operators with an ${\cal R}$-charge of $J=O(\sqrt{N})$. These objects are already heavy enough that, for single
traces we have a new effective string coupling replacing ${1\over N}$
$$ g_s\sim {1\over N}\to {J^2\over N}\, .$$
Thus, to suppress non-planar corrections we need to take $J^2\ll N$, which we assume from now on. To see stringy physics it is useful
to consider operators, for example, of the form
$$ \Tr (YZ^n YZ^{J-n})\, .$$
Due to the presence of the $Y$ fields this state has $\Delta =J+2$ and ${\cal R}$-charge $J$. Since $J^2\to\infty$, this operator
is nearly ${1\over 2}$-BPS and we might still expect that corrections are suppressed. This is indeed the case \cite{Berenstein:2002jq}:
one finds that the expansion parameter $\lambda$ is replaced by ${\lambda\over J^2}$. The eigenvalues of the dilatation operator
when acting on this class of states precisely matches the expected energies of strings in the dual string theory\cite{Berenstein:2002jq}. 
The matching of
spectra can be significantly improved. Think of the $Y$s and $Z$s as populating a 
lattice with $J+2$ sites. Further, identify the $Y$s with a spin up state and the $Z$s with a spin down state. With this
interpretation, the Yang-Mills dilatation operator can be identified with the Hamiltonian of a spin 
chain\cite{Minahan:2002ve,Beisert:2003yb,Beisert:2003tq}. By considering coherent states of the spin chain,
one can give the spin chain a sigma model description. The resulting sigma model agrees exactly with the sigma model for a 
string rotating with a large angular momentum, so that now agreement is obtained at the level of the action\cite{Kruczenski:2003gt}.
The very detailed agreement allows one to frame very precise questions. For example, it is straight forward to show that the 
mean value of the spin of the spin chain corresponds to the position of the string in the dual gravitational spacetime (see
\cite{Kruczenski:2003gt} for details). 

{\vskip 0.5cm}

\noindent
{\sl {\bf Key idea:} Operators with an ${\cal R}$-charge of $O(\sqrt{N})$ do indeed correspond to strings. The stringy dynamics
can be recovered from the field theory and further, it is clear how to build field theory states corresponding to strings localized 
at a point in the dual (higher dimensional!) gravitational theory.}

\section{Giant Gravitons}

In this section we study certain membranes in string theory, known as giant gravitons. These operators have an ${\cal R}$-charge of $O(N)$; 
in this case the trace basis badly fails to provide orthogonal states and hence we have the first case in which the Schur polynomials {\it must} 
be used. 

Giant graviton solutions describe branes extended in the sphere\cite{McGreevy:2000cw} or in the 
AdS space \cite{Grisaru:2000zn,Hashimoto:2000zp},
of the AdS$\times$S background. The giant gravitons are (classically) stable\footnote{They
carry an RR-dipole charge, but no RR-monopole charge. Thus their decay would not be prevented by
charge conservation.}
due to the presence of the five form flux which produces a force that exactly
balances their tension. The force which balances the tension is a Lorentz-like\footnote{i.e. velocity
dependent} force so that the force increases with increased giant angular momentum. Consequently
a giant graviton expands to a radius proportional to the square root of its 
angular momentum\cite{McGreevy:2000cw}. If the giant is expanding in the S$^5$ of the
AdS$\times$S background, there is a limit on how large it can be - its radius must be less than the 
radius of the S$^5$\cite{McGreevy:2000cw}. This in turn implies a cut off on the angular momentum of the giant. Since angular momentum 
of the giant maps into ${\cal R}$-charge, there should be a cut off on the ${\cal R}$-charge 
of the dual operators. 
The Schur polynomials corresponding to totally antisymmetric representations do have a cut off on their
${\cal R}$-charge; this cut off exactly matches the cut off on the giant's angular momentum\cite{Balasubramanian:2001nh}. 
Thus, it is natural to identify Schur polynomials for the completely antisymmetric representations as the operator
dual to sphere giant gravitons. Another class of Schur polynomials which are naturally singled out, are those corresponding to
totally symmetric representations. Since these representations are not cut off, they are naturally identified as operators
dual to AdS giant gravitons\cite{Corley:2001zk}, which, because they expand in the AdS space, can expand to an arbitrarily 
large size and hence have no bound on their angular momentum. The Young diagram has at most 
$N$ rows, implying a cut off on the number of AdS giant
gravitons; the need for this cut off is again visible in the dual gravitational theory: it ensures
that the five form flux at the center of the AdS space does not become zero; a non-zero flux is needed to support an AdS giant. 

A smoking gun signal for D-branes in string theory, is the presence of open strings in the excitation spectrum of the D-brane.
The low energy dynamics of these strings should realize a new gauge theory on the brane worldvolume. Since the brane worldvolume
is a compact space, Gauss' Law will only allow excitations with vanishing total charge. There is a very natural generalization
of the Schur polynomials, the restricted Schur polynomials, in which the symmetric group character appearing in (\ref{Schur})
is replaced by a partial trace over the symmetric group elements. The number of partial traces that can be defined matches
the number of states consistent with the Gauss Law constraint\cite{Balasubramanian:2004nb}.
 \begin{figure}[h]
          \centering
          {\epsfig{file=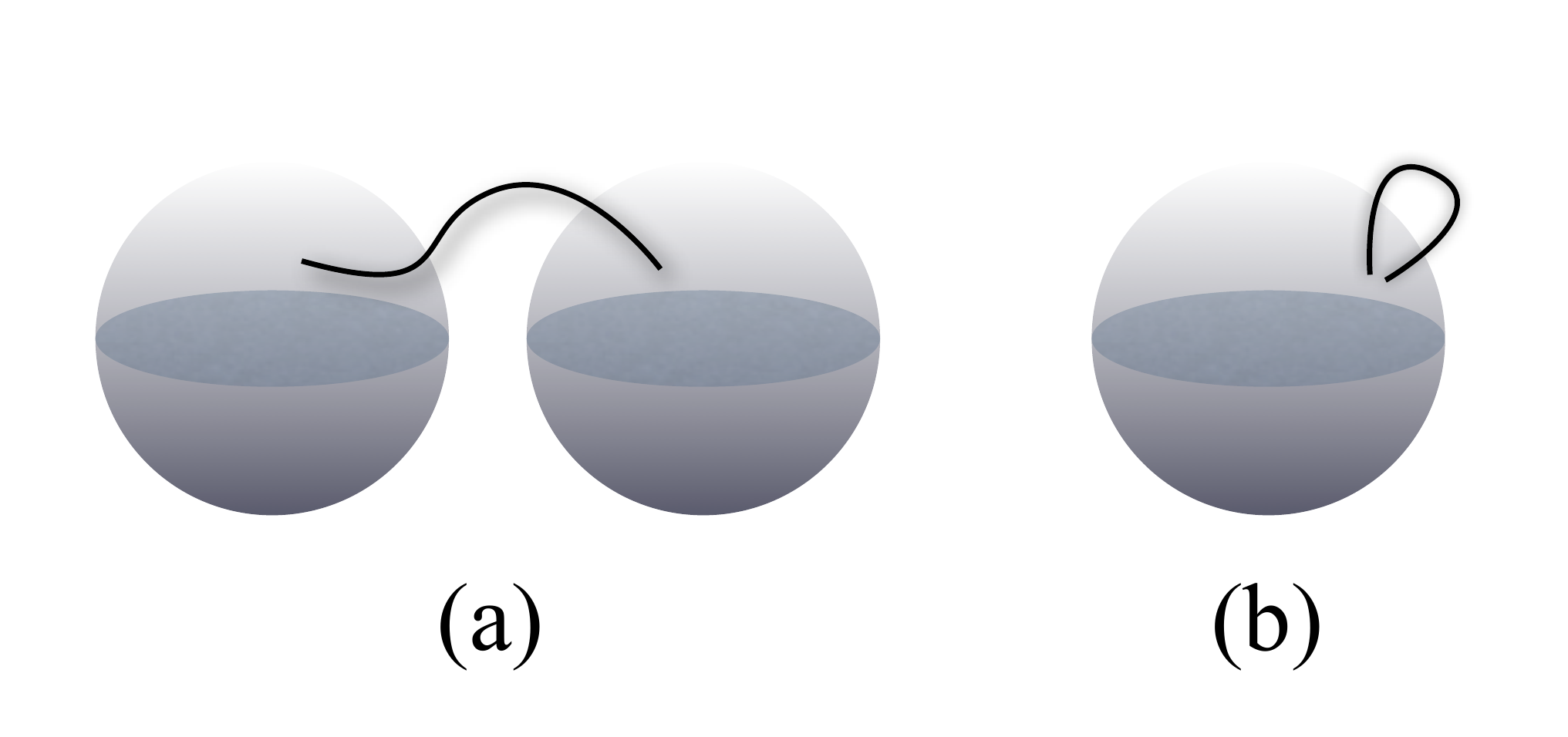,width=7.5cm,height=3.5cm}}
          \caption{Giants with strings attached}
 \end{figure} 
Surprisingly, it is still possible
to compute correlators of restricted Schur polynomials and to evaluate the action of the field theory dilatation operator
on these operators\cite{de Mello Koch:2007uu,Berenstein:2006qk,de Mello Koch:2007uv}. This is rather non-trivial: due to
the large ${\cal R}$-charge of the operators one
needs to sum an infinite number of non-planar diagrams in these computations. 
The dilatation operator can again be matched to the action for open strings, 
at the level of the action\cite{Berenstein:2006qk}.
This allows one to ask questions that could not be asked in perturbative string theory. 
For example, when the string and membrane interact, the Young diagram changes shape. This allows one to take
backreaction on the membrane into account\cite{Berenstein:2006qk,de Mello Koch:2007uv}. From the point of view of perturbative
string theory, the D-brane appears as a boundary condition and it is not obvious how one should account for backreaction.
Another interesting effect discovered by studying these operators is an instability arising when long open strings are
attached to the giant graviton\cite{Berenstein:2006qk}. The giant graviton, which couples to the RR-five form flux, does
not undergo geodesic motion. The open string, which does not couple to the RR-five form flux, would like to undergo
geodesic motion but is being dragged in a non-geodesic motion by the giant. These centrifugal forces can overcome the
string tension if the string is long enough, leading to the instability. This effect has also been exploited as a toy
model for quantum gravity effects in braneworld cosmological models\cite{jeff}.

\begin{center}
  \fbox{
    \begin{minipage}[c]{14cm}
      \small{
        {\vskip 0.1cm}
        \center{\bf Emergent Gauge Theory}\\
        {\vskip 0.01cm}
        \noindent
        \flushleft{Fundamental strings are charged under the Kalb-Ramond 2-form. 
        For stretched strings this
        charge can be thought of as a current flowing along the string. Consequently, when they
        end on D-branes, conservation of string charge means that strings act as `sources' or
        `sinks' on the brane worldvolume. When $-$ as in the case of the giant gravitons above 
        $-$ the D-brane wraps a compact space like the $S^{3}\subset S^{5}$, 
        the only allowed attached open string states are those consistent with Gauss' law. So, for 
        example, for the giant configurations in Figure 1. only (b) is a valid state. This is a 
        manifestation of the closed topology of the spherical D-brane.
       To see how Gauss' law is encoded in the Yang-Mills
        theory we need to consider operators dual to {\it excited} giant gravitons.
        These {\it restricted} Schur polynomials,} 
        \begin{eqnarray}
          \chi^{(k)}_{R,R_{1}} = \frac{1}{(n-k)!}\sum_{\sigma\in S_{n}}\mathrm{Tr}_{R_{1}}
          \left(\Gamma_{R}\left(\sigma\right)\right)\mathrm{Tr}\left(\sigma 
          Z^{\otimes n-k}\left(W^{(1)}\right)^{i_{n-k+1}}_{i_{\sigma(n-k+1)}}\cdots
          \left(W^{(k)}\right)^{i_{n}}_{i_{\sigma(n)}}\right),\nonumber
        \end{eqnarray}
        \flushleft{are obtained from (\ref{Schur}) by the insertion 
        of `words' $W^{(i)}$ describing the open strings attached to the giant system. 
        In the language of Young diagrams, this corresponds to labeling boxes in the 
        diagram associated to the giant graviton. 
        For the single string attached to single sphere giant graviton with momentum $p$ in 
        Fig 1.(b), 
        for example, the restricted Schur takes the form 
        $\displaystyle \chi^{(1)}_{{\bf 1}^{p+1},{\bf 1}^{p}}(Z,W)$ with the open string 
        word $W^{i}_{j} = (Y^{J})^{i}_{j}$ say. We have denoted a Young diagram with a 
        single column
        containing $n$ boxes by $1^n$. To match with the gravity side of the 
        correspondence, we need to count the number of possible operators  
        $\chi^{(k)}_{R,R_{1}}(Z,W^{i})$ that can be constructed for a given representation. 
        Remarkably, the number of these 
        operators matches precisely with the number of allowed states in the string theory 
        that satisfy the Gauss constraint\cite{Balasubramanian:2004nb}.}
        } 
    \end{minipage}
  }
\end{center}
{\vskip 0.5cm}
\noindent
{\sl {\bf Key idea:} Giant gravitons are membrane states that are represented in the dual field theory by Schur polynomials and 
restricted Schur polynomials. The field theory correctly reproduce both the number of open string excitations of the giant gravitons
and their dynamics.}

\section{New Geometries}

When the operator we consider has an ${\cal R}$-charge of $O(N^2)$, we are studying objects in the dual string theory that are so heavy
that they backreact on spacetime. Consequently, there should be new supergravity backgrounds corresponding to these operators. These
new backgrounds should preserve an $R\times SO(4)\times SO(4)$ symmetry. The conserved charge associated to $R$ corresponds to
conformal dimension in the field theory. The first $SO(4)$ symmetry corresponds to rotations of the $S^3$ on which the field theory
is defined (recall that our operators are built from the $s$-wave of $Z$). The second $SO(4)$ corresponds to ${\cal R}$-symmetry rotations
of the four scalars $\phi_3,\phi_4,\phi_5,\phi_6$ which are not used to construct $Z$. This ansatz is sufficiently specific that the general
solution with these isometries can be written down. These geometries, constructed by Lin, Lunin and Maldacena, are known as the
LLM geometries\cite{Lin:2004nb}. The complete solution is determined in terms of a single function which obeys a three dimensional
Laplace equation. To get a unique solution, one needs to specify a boundary 
condition for this function on a specific two dimensional plane.
In order that the geometry is regular, this boundary condition must assign the function either of the values $\pm{1\over 2}$ on this plane.
It is tempting to identify this two dimensional plane as the phase space of the fermions of section 2, where regions with ${1\over 2}$
correspond to occupied states and regions with $-{1\over 2}$ correspond to unoccupied states\cite{Lin:2004nb}. 
This can indeed be checked in detailed computations and it turns out that this identification is perfect\cite{Grant:2005qc}. 
Thus, once again the 
quantum field theory and the quantum gravity theories are in complete agreement.

Denote the Schur polynomial of ${\cal R}$-charge of $O(N^2)$ by $\chi_B (Z)$. The operator $\chi_B (Z)$ creates the new background.
Is it possible, directly in the field theory, to construct the metric corresponding to this dual geometry?
One thing we could do is follow the propagation of a graviton in the spacetime. 
Since we know how to build an operator dual to a graviton, this is a computation we know how to do. 
Further, gravitons move along null geodesics and hence they ``know'' about the dual geometry, 
so that we might indeed hope to learn something about the metric.
To create a graviton moving on the new background we need to act with $\Tr (Z^n)$ with $n$ a number of $O(1)$. 
Thus, to probe the geometry we need to compute
the correlator
$$ \left\langle \chi_B(Z)\chi_B(Z)^\dagger\Tr (Z^n)\Tr (Z^n)^\dagger\right\rangle\, .$$
This computation can be performed exactly, in the free field limit, for any choice of the Young diagram 
$B$\cite{Koch:2008ah,Koch:2008cm}. As soon as $B$ has $O(N^2)$ boxes, one can no longer neglect Wick contractions
between fields in $\Tr (Z^n)^\dagger$ and fields in the background $\chi_B(Z)$ - at precisely the value of the
${\cal R}$-charge that we expect back reaction is important, the graviton and background start interacting! This
is how the field theory accounts for back reaction\cite{Koch:2008ah,Koch:2008cm}. 
The 
resulting correlators are surprisingly simple. If one takes
$B$ to be a Young diagram with $M$ columns ($M$ is $O(N)$), 
the net effect of the background is to replace $N\to M+N$.
Recall that in the trivial background graviton correlators admit a ${1\over N}$ expansion.
In the new background, graviton correlators are organized by a ${1\over M+N}$ expansion\cite{Koch:2008ah,Koch:2008cm}. This
renormalization of the string coupling constant was achieved by summing an infinite number of non-planar
diagrams, something that is only possible thanks to the power of the Schur polynomial technology.
This renormalization of the string coupling can be checked rather explicitly, using holography in the LLM
background\cite{Skenderis:2007yb}, along the lines of the computation of section 5. 
One again finds perfect agreement\cite{Koch:2009jc}. Apart from probing the geometry with gravitons, one could
consider probing the geometry with strings\cite{Vazquez:2006id,Chen:2007gh} or even giant gravitons\cite{Mandal:2005wv}.
This leads to some interesting results. For example, in the case of string probes, one can again construct the dilatation
operator and study the sigma model that arises from the coherent state expectation value of the model. In this
way it is possible to read off the metric that the string feels\footnote{The string only moves close to the
plane on which the boundary condition is specified. One is able to read off the metric on this plane which is what
we mean by ``the metric that the string feels''. Of course one would like to do better and determine
the full metric.}\cite{Vazquez:2006id,Chen:2007gh}, which is rather
detailed information about the dual geometry.

Given this very concrete description of the ${1\over 2}$-BPS geometries, it is possible to reexamine
some long standing puzzles. An important problem in this class is the information loss paradox.
The entropy of black holes suggests an enormous degeneracy of microscopic states. The information loss
paradox would be evaded if one could show that a pure initial state collapses to a particular
pure black hole microstate whose exact structure can be deduced by careful measurements. 
What do pure microstates look like and what sorts of measurements can distinguish them
from each other? In \cite{Balasubramanian:2005mg}, this problem was examined by applying information
theoretic ideas to Schur polynomials with $O(N^2)$ boxes. It is possible to characterize a ``typical
operator'' and then to ask what the semi-classical description, in the dual gravity, of this state is.
Concretely, the half-BPS sector is integrable and the eigenvalues of the resulting set of commuting conserved charges completely identify a state. 
These charges can be measured in the dual gravitational description from asymptotic multipole moments of the spacetime.  
A typical very heavy state corresponds to a spacetime ``foam''. 
Almost no semiclassical probes will be able to distinguish different foam states\cite{Balasubramanian:2006jt},
and the resulting effective description gives a singular geometry\cite{Balasubramanian:2007zt}.
Although the ${1\over 2}$-BPS states considered are not black holes,
this study seems to explain how the existence of pure underlying
microstates and the absence of fundamental information loss are consistent with the
thermodynamic character of semiclassical black holes\cite{Balasubramanian:2005mg}.
\noindent
\begin{center}
  \fbox{
    \begin{minipage}[c]{14cm}
      \small{
        {\vskip 0.05cm}
        \center{\bf Local Gravitons}\\
        \noindent
        \flushleft{For nearly a century now, we have learnt that to understand the 
        geometry of spacetime, we have to probe it with localised objects; (very roughly) 
        if you want to know 
        whether a spacetime is curved or not, throw some particles into it and chart 
        their geodesics. From a field-theorist's point of view, local geometric structure 
        in spacetime arises as a coherent excitation of gravitons. To understand how 
        bulk spacetime geometry {\it emerges} in the Yang-Mills matrix model, one might
        try to understand the localization and dynamics of gravitons.}\\
        
        \flushleft{
        To this end, following the arguments outlined above, the normalized operator dual to a
        graviton of one unit of angular momentum (in $S^{5}$) is}
        $\displaystyle \frac{\mathrm{Tr}(Z)}{\sqrt{N}}$. Identifying this operator with 
        the graviton creation operator $a^{\dagger}$ in the dual quantum gravity theory, we can 
        also define a graviton annihilation operator $\displaystyle \frac{1}{\sqrt{N}}
        \mathrm{Tr}\left(\frac{d}{dZ}\right) \leftrightarrow a$, so that $\left[a,a^{\dagger}
        \right] = 1$. From this, a graviton coherent state operator can be built in the usual way,
        \begin{eqnarray}
           \mathcal{O}_{z} = e^{-\frac{1}{2}|z|^{2}}e^{z\frac{\mathrm{Tr}(Z)}{\sqrt{N}}}\, .
           \nonumber
        \end{eqnarray}}
        \flushleft{After conformal mapping to $R\times S^{3}$, the operator 
        $\mathcal{O}_{z}$ is mapped into the coherent state $| z\rangle$ with $z = 
        r\,e^{-i\phi}$. The dynamics of the low energy excitations of this coherent state
        on $R\times S^{3}$ is captured by the Landau-Lifshitz Lagrangian}
        \begin{eqnarray}
          L = \langle z| i\frac{d}{dt}|z\rangle - \langle z| H|z\rangle = \dot{\phi}r^{2} - r^{2},
          \nonumber
        \end{eqnarray}
        \flushleft{when evaluated on the graviton coherent state. 
        The equations of motion can be integrated
        to determine $\dot{\phi} = 1$ and $\dot{r} = 0$, which are, by now familiar results for 
        gravitons in $\mathrm{AdS}_{5}\times \mathrm{S}^{5}$. These results are not unique
        to $\mathrm{AdS}_{5}\times \mathrm{S}^{5}$ and can be extended to a whole class 
        of $\frac{1}{2}-$BPS geometries - the so-called LLM backgrounds which are built from 
        operators with $\mathcal{R}-$charge of $\mathcal{O}(N^{2})$. These results can be extended
        to describe gravitons localized at different values of $r$ in the multi ring LLM geometries 
        providing compelling evidence 
        that {\it local} geometry emerges from the super Yang-Mills gauge theory\cite{Koch:2008ah}.}
    \end{minipage}
  }
\end{center}

{\vskip 0.5cm}

\noindent
{\sl {\bf Key idea:} The field theory description of ${1\over 2}$-BPS geometries is in terms of operators with dimension of $O(N^2)$.
Probing this geometry corresponds to computing correlators of operators with the background operator inserted. The correlators can be used
to extract rather detailed information about the dual geometry.}

\section{Outlook}

It is clear that something non-trivial is working. Things are however, far from satisfactory. 
The AdS/CFT correspondence has passed highly non-trivial tests 
(far more than we discussed) so that we are confident the
basic idea is correct. However, we still have no real understanding of why the degrees of freedom of a 
strongly coupled Yang-Mills
theory are most simply described starting from a higher dimensional dynamical geometry.
The geometry is not visible in the weak coupling
(Lagrangian) description of the quantum field theory, and in this sense is emergent.
A simple example of an emergent geometry is provided by the large $N$ limit of
a single Hermitian matrix model quantum mechanics\cite{Das:1990kaa}.
In the large $N$ limit, the integral is dominated by a saddle point with a definite
eigenvalue distribution. The emergent geometry relies crucially on the repulsive 
inter eigenvalue force which causes the eigenvalues to spread. Without this repulsive
force the eigenvalues would simply sit at one of the minima of the potential of the matrix model.
The repulsive force itself comes from the Van der Monde determinant, that is, from the measure
of the path integral\footnote{The Van der Monde determinant has recently been shown to arise
in certain sectors of multimatrix models\cite{JS}.}. 
It is thus a quantum effect, which seems to match nicely with the fact that
$\hbar$ sets the radius of the AdS space (see footnote 1 on page 2). To go beyond a single
matrix is extremely difficult. For a single matrix model, the eigenvalues provide a very convenient
set of variables. For more than one matrix, since the matrices will not in general commute,
its not clear what the analog of the eigenvalues are. An important fact might be that we
only expect the emergence of geometry in the strong coupling limit.
Berenstein\cite{Berenstein,Berenstein2} has suggested that in the strong coupling and large $N$ limit 
the matrices will commute with each other. In this way the usual
$N^2$ degrees of freedom of matrices get effectively reduced to order $N$ degrees of freedom
and the collective description of these low energy degrees of freedom can
be given in terms of a joint eigenvalue distribution for several matrices. It is the
geometrical description of this eigenvalue distribution that is supposed to produce the emergent 
geometry\cite{Berenstein,Berenstein2}.

The emergent geometry obtained in the large $N$ limit of
a single Hermitian matrix model quantum mechanics\cite{Das:1990kaa}
was constructed by making a systematic change of variables in the quantum
theory, correctly accounting for the Jacobian of this change. A systematic
way to achieve this is provided by the collective field theory
formalism\cite{cft}. For a single matrix a convenient set of gauge invariant
variables is provided by traces of powers of the matrix. For more than one matrix
one needs to consider traces of arbitrary words built from the matrices - which is
rather complicated. However, such a rewriting does indeed show that the field
theory reconstructs the dual gravitational dynamics\cite{friends}. In principle this approach
would provide a way to systematically reconstruct the dual gravitational dynamics
from the field theory. The problem however, is one of considerable complexity.
It seems that additional insight is needed before this program can be taken to completion.

Above we have restricted ourselves to the ${1\over 2}$-BPS sector. A skeptic might suggest
that it is dangerous to draw general lessons from the ${1\over 2}$-BPS sector.
Indeed it might be, so we should try to do better.
To go beyond the half-BPS sector, one needs to study multi-matrix dynamics. 
In general, this is a formidable problem.
There has however, been some recent progress: three independent bases for general multi-matrix models have been identified.
For a review of these developments and the work leading up to them, see \cite{sanjaye}.
The  basis described in \cite{Brown:2007xh} builds operators with definite flavor quantum numbers. 
The basis of \cite{Kimura:2007wy} uses the Brauer algebra to build correlators involving $Z$ and $Z^\dagger$;
this basis seems to be the most natural for exploring brane/anti-brane systems.
The basis of \cite{Bhattacharyya:2008rb} most directly allows one to consider excitations of the operator.
All three bases diagonalize the two point functions in the free field theory limit (to all orders in
the ${1\over N}$ expansion); it is in this sense that they generalize the Schur polynomials to the case of
multimatrix models. The discovery of these bases seems to be a promising start towards exploring 
${\cal N}=4$ super Yang-Mills theory beyond the ${1\over 2}$-BPS
sector. On the gravity side there has been some progress in characterizing the ${1\over 4}$-BPS
geometries\cite{Lunin:2008tf}.

Perhaps in the not too distant future we will have an answer for George.

{\vskip 0.5cm}

\noindent
{\it Acknowledgements:} 
We would like to thank Rajsekhar Bhattacharyya, Tom Brown, Storm Collins, Tanay Dey,
George Ellis, Alex Hamilton, Norman Ives, Antal Jevicki, Yusuke Kimura,
Andrea Prinsloo, Sanjaye Ramgoolam, Joao Rodrigues, Jelena Smolic, 
Milena Smolic, Michael Stephanou, Dave Turton,
Nick Toumbas and Alex Welte for pleasant discussions and/or helpful correspondence. 
This work is based upon research supported by the South African Research Chairs
Initiative of the Department of Science and Technology and National Research Foundation. JM is supported by the National Research Foundation under the Thuthuka and Key International Scientific Collaboration programs.
Any opinion, findings and conclusions or recommendations expressed in this material
are those of the authors and therefore the NRF and DST do not accept any liability
with regard thereto.

\end{document}